# Estimation of Pelvic Sagittal Inclination from Anteroposterior Radiograph Using Convolutional Neural Networks: Proof-of-Concept Study

Ata Jodeiri[1,2]\*, Yoshito Otake[2], Reza A. Zoroofi[1], Yuta Hiasa[2], Masaki Takao[3], Keisuke Uemura[3], Nobuhiko Sugano[3], Yoshinobu Sato[2]

[1] School of Electrical & Computer Engineering, University of Tehran, Tehran, Iran

[2] Graduate School of Information Science, Nara Institute of Science and Technology, Nara, Japan

[3] Department of Orthopaedic Surgery, Osaka University Graduate School of Medicine, Suita, Japan

## INTRODUCTION

Alignment of the bones in standing position provides useful information in surgical planning. In total hip arthroplasty (THA), pelvic sagittal inclination (PSI) angle in the standing position is an important factor in planning of cup alignment [1] and has been estimated mainly from radiographs. Previous methods for PSI estimation [2], [3] used a patient-specific CT to create digitally reconstructed radiographs (DRRs) and compare them with the radiograph to estimate relative position between the pelvis and the x-ray detector. In this study, we developed a method that estimates PSI angle from a single anteroposterior radiograph using two convolutional neural networks (CNNs) without requiring the patient-specific CT, which reduces radiation exposure of the patient and opens up the possibility of application in a larger number of hospitals where CT is not acquired in a routine protocol.

## MATERIALS AND METHODS

As shown in Figure 1, our approach consists of two main parts including CNNs for the segmentation and for the PSI estimation. In the following, we describe the dataset preparation and details of the architecture of the segmentation network and PSI estimation network, respectively. In this preliminary study, we evaluated each network separately.

In this study, we used a database of 472 CTs acquired for the planning of total hip replacement surgery. First we automatically segmented the pelvis in all CTs using a previously developed method [4]. The landmarks defining PSI, namely anterior superior iliac spines (ASIS) and pubic tubercles (PT) of both sides, were identified by an expert surgeon manually for all CTs. Our training dataset detailed below is created from this 472 pairs of CT, mask image of the pelvis, and the anatomical landmarks.

The purpose of the segmentation network is to separate the shape of the pelvis from the radiograph. We used the U-net architecture [5] with 19 3×3 convolutional layers, 4 2×2 max-pooling layers, 4 2×2 up-convolution layer and 4 copy and crop parts as shown in Figure 1. The network was trained by DRRs and corresponding pelvis mask (i.e., the mask in CT projected onto the virtual detector plane). In the evaluation experiment, we created 1000 DRRs from 15 CTs (i.e., 15,000 DRRs in total) with random transformation (±15 [deg] rotation and ±20 [mm] translation). We tested the segmentation network with 100 DRRs that were created from 100 CTs, which was not included in the training dataset, using random transformations (±15 [deg] rotation and ±20 [mm] translation).

Following the segmentation network, we extract the pelvis region from the radiograph and apply another CNN to estimate the PSI. The PSI estimation network is also trained by DRRs. It performs regression using a network consisting of 6 layers include three 5×5 convolutional layers with stride 2 and three fully-connected layers with Rectified Linear Unit (ReLU) activation function. The first and second fully-connected layers include 100 and 250 activations neurons, respectively. The training dataset of PSI estimation network includes 500 DRRs for 472 patients with a random transformation (±15 [deg] rotation in anteroposterior direction and ±5 [deg] rotation in two other directions and ±20 [mm] translation). The PSI corresponding for each DRR was computed by the landmarks manually identified in the CT. For evaluation of the PSI estimation network, we performed 4-fold cross validation and calculated the error in PSI.

The experiments were conducted on a workstation with Intel Xeon(R) and 2.8 GHz CPU with 32GB RAM and Nvidia GeForce GTX 980 GPU. Both CNNs were implemented with cuDNN acceleration using a deep learning framework, Chainer (https://chainer.org/).

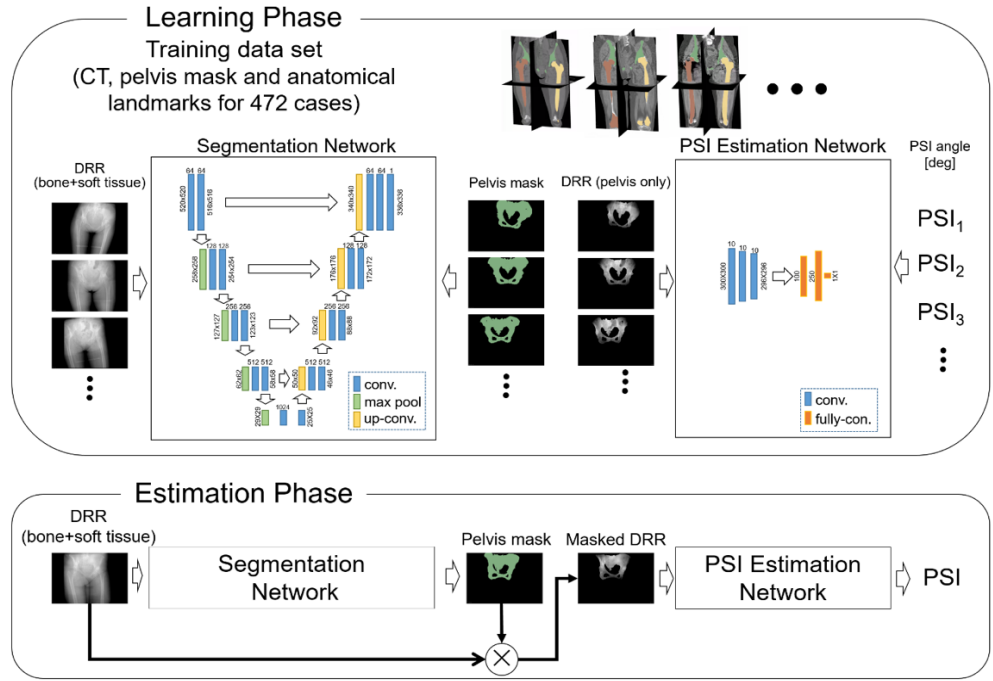

**Figure 1:** Overview of the proposed method, which includes two CNNs, one for segmentation to extract pelvis shape in the projection image and the other for estimation of PSI from the projected pelvis shape.

## RESULTS

Figure 2(a) shows the Dice coefficient for the 100 test DRRs. The mean and standard deviation for Dice coefficient were 0.947 and 0.019 that suggested an accurate segmentation with low deviation. Figure 2(b) shows the boxplot of PSI errors in 472 patients. The accuracy was dependent on the shape of the pelvis. In Figure 2(c) the upper DRR shows a pelvis that yielded larger error in PSI estimation, while the pelvis shape in the lower DRR yielded a high accuracy for any angles. Figure 2(d) shows a scatter plot of estimated PSI and the ground truth PSI for all patients (we show the results of only 10 DRRs for every patient to improve visibility of the plot). The PSI estimation error for all cases was 3.22 ± 2.18 [deg].

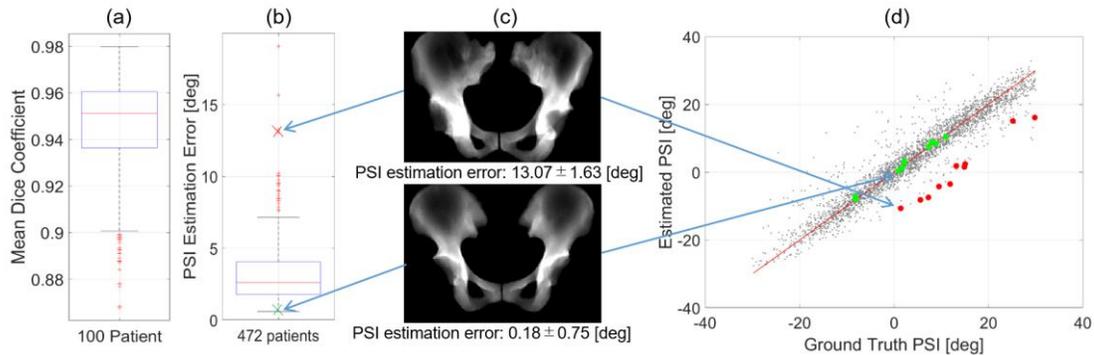

**Figure 2:** Results of the evaluation experiments. (a) boxplot of mean Dice coefficient for segmentation on 100 patients. (b) boxplot of PSI estimation for 472 patients. (c) two example DRRs, upper: the shape yielded larger errors, lower: the shape yielded small errors (d) scatter plot of the estimated PSI and the ground truth PSI for 472 (only 10 DRRs for every case were shown for improving visibility), red large dots indicate the results corresponding to the upper pelvis shape and green indicate the results corresponding to the lower pelvis shape.

## DISCUSSION

Pelvic tilt that has been measured by PSI affects functional anteversion and inclination of acetabular components and estimation of that would be very important before THA for avoiding of implant dislocation. We proposed a method to predict PSI directly from a single 2D hip radiograph. It can be applied in many clinical setups in which we do not have access to the patient-specific CT. In order to predict the PSI value, we introduced 2 CNNs, one for segmentation of the pelvis from the radiograph and second for the estimation of PSI from the radiograph including only the pelvis region.

Through the quantitative evaluation experiments, we found that for 75% of the patients (354 cases) the PSI estimation error was relatively small ($2.24 \pm 0.78$ [deg]) for any PSI angles, while the other 25% of the patients (118 cases) resulted in a larger error ($6.14 \pm 2.38$ [deg]). Based on visual assessment of the DRRs, We believe the large error in 25% of patients is due to the inaccurate segmentation of pelvis in CT images.

Our future work includes tuning of the both network architectures and the hyper parameters to improve the estimation performance. Also, estimation of the pelvis shape (e.g., location of the anatomical landmarks in 3D) from radiograph is underway.